\documentclass{article}
\usepackage{amssymb}
\usepackage{amsmath}

\newtheorem{theorem}{Theorem}
\newtheorem{acknowledgement}[theorem]{Acknowledgement}

\input{tcilatex}
\begin{document}

\title{Is the Schwarzschild Metric a Vacuum Solution of the Einstein
Equation?}
\author{Horace Crater\thanks{%
hcrater@utsi.edu}~ \\
%EndAName
The University of Tennessee Space Institute}
\maketitle

\begin{abstract}
\ This paper examines the inhomogeneous Einstein equation for a static
spherically symmetric metric with a source term corresponding to a\ perfect
fluid with $p=-\rho $.\ By a careful treatment of the equation near the
origin we find an analytic solution for the metric, dependent on a small
parameter $\varepsilon ,$ which can be made arbitrarily close to the
Schwarzschild solution as $\varepsilon \rightarrow 0$ and which in that same
limit can be viewed as arising from a point-like source structure. \ 
\end{abstract}

\section{\protect\bigskip Introduction}

This paper examines solutions of the inhomogeneous Einstein equation 
\begin{equation}
G_{\mu \nu }=R_{\mu \nu }-\frac{1}{2}Rg_{\mu \nu }=-8\pi GT_{\mu \nu }\equiv
-\kappa T_{\mu \nu },  \label{ee}
\end{equation}%
for static spherically symmetric metrics corresponding to a perfect fluid at
rest with $p=-\rho $. \ Our aim is to develop an analytic solution for the
metric that can be made arbitrarily close to the Schwarzschild solution but
yet retains a nonvanishing contribution to the source term, unlike the
Schwarzschild solution. At the same time we find that the second solution
with the same source term is arbitrarily close to the de Sitter solution.
The Einstein equations for the metric are second order and highly nonlinear.
\ This implies that if one has two independent solutions, then their linear
combination will not be a solution. Nevertheless, as we emphasize in the
first section, under those special circumstances for the source term in the
Einstein equation, the metric can be written in terms of a potential $\phi ~$%
which satisfies a very simple \textit{linear} homogeneous second order
differential equation, with two linearly independent solutions. One of the
potential-like functions ($\sim 1/r$) is the Newtonian gravitational
potential which we associate with the Schwarzschild solution while the other
($\sim r^{2}$) is a potential associated with the de Sitter solution
(together they compose what are known as the de Sitter Schwarzschild-
solution \cite{desch}, \cite{desich},\cite{rin}). \ The contribution of the
latter to the source term is a constant while that of the Schwarzschild
solution has a vanishing contribution. In order to accomplish our aim, we
recast the inhomogeneous Einstein equation as the limiting case of a closely
related inhomogeneous equation dependent on a small parameter $\varepsilon $%
. \ A careful treatment of this problem near the origin leads to source
terms with two separate non-vanishing contributions

\section{Solutions of the Inhomogeneous Einstein Equation}

\subsection{Static Spherically Symmetric Solutions of the Inhomogeneous
Einstein Equation\ for Perfect Fluid with $p=-\protect\rho .$}

In this section we first review the static spherically symmetric standard
solution of the vacuum Einstein equation $G_{\mu \nu }=0$. \ The equation is
second order in the above sense and has two solutions with one being the
Schwarzschild solution while the other a constant. \ We then remind the
reader how if one adds a source term corresponding to a perfect fluid at
rest with $p=-\rho $ \ a second nonconstant solution emerges in addition to
the Schwarzschild one. \ The pressure and density are found to be constants
and the second solution is one found originally by de Sitter for the
Einstein equation with a cosmological constant\cite{desit2},\cite{desit4}
and is a static form of the \ time dependent one used in models of inflation
and dark energy in modern cosmology. \ 

For a spherically symmetric solution one chooses the coordinates 
\begin{eqnarray}
x^{0} &=&t,  \notag \\
x^{1} &=&r,  \notag \\
x^{2} &=&\theta ,  \notag \\
x^{3} &=&\phi .
\end{eqnarray}%
In a vacuum with static conditions as well as spherical symmetry, we use
Dirac's exponential parametrization of the metric,\cite{dirac}, 
\begin{eqnarray}
d\tau ^{2} &=&e^{2\nu (r)}dt^{2}-e^{2\lambda (r)}dr^{2}-r^{2}(d\theta
^{2}+\sin ^{2}\theta d\phi ^{2}),  \notag \\
g_{00} &=&-e^{2\nu (r)}=1/g^{00}  \notag \\
g_{11} &=&e^{2\lambda (r)}=1/g^{11},  \notag \\
g_{22} &=&r^{2}=1/g^{22},  \notag \\
g_{33} &=&r^{2}\sin ^{2}\theta =1/g^{33}.  \notag \\
g_{\mu \nu } &=&0,~\mu \neq \nu .  \label{1}
\end{eqnarray}%
With 
\begin{equation}
\Gamma _{\mu \nu }^{\kappa }=\frac{g^{\kappa \sigma }}{2}(g_{\nu \sigma ,\mu
}+g_{\mu \sigma ,\nu }-g_{\mu \nu ,\sigma })=\Gamma _{\nu \mu }^{\kappa },
\end{equation}%
the only nonzero $\Gamma ^{\prime }s$ are \cite{dirac}%
\begin{eqnarray}
\Gamma _{00}^{1} &=&\nu ^{\prime }e^{2\nu -2\lambda },~~~~~~~~\Gamma
_{10}^{0}=\nu ^{\prime },  \notag \\
\Gamma _{11}^{1} &=&\lambda ^{\prime },~~~~~~~~~~~~~~~~~\Gamma
_{12}^{2}=\Gamma _{13}^{3}=r^{-1},  \notag \\
\Gamma _{22}^{1} &=&-re^{-2\lambda },~~~~~~~~~~\Gamma _{23}^{3}=\cot \theta ,
\notag \\
\Gamma _{33}^{1} &=&-r\sin ^{2}\theta e^{-2\lambda },~~\Gamma
_{33}^{2}=-\sin \theta \cos \theta .  \label{afin}
\end{eqnarray}%
With%
\begin{equation}
R_{\nu \sigma }=\Gamma _{\nu \lambda }^{\lambda },_{\sigma }-\Gamma _{\nu
\sigma }^{\lambda },_{\lambda }+\Gamma _{\nu \sigma }^{\kappa }\Gamma
_{\kappa \lambda }^{\lambda }-\Gamma _{\nu \kappa }^{\kappa }\Gamma _{\sigma
\lambda }^{\lambda },
\end{equation}%
the diagonal elements of the Ricci tensor are%
\begin{eqnarray}
R_{00} &=&\left( -\nu ^{\prime \prime }+\lambda ^{\prime }\nu ^{\prime }-\nu
^{\prime 2}-\frac{2\nu ^{\prime }}{r}\right) e^{2\nu -2\lambda },  \notag \\
R_{11} &=&\nu ^{\prime \prime }-\lambda ^{\prime }\nu ^{\prime }+\nu
^{\prime 2}-\frac{2\lambda ^{\prime }}{r},  \notag \\
R_{22} &=&(1+r\nu ^{\prime }-r\lambda ^{\prime })e^{-2\lambda }-1,  \notag \\
R_{33} &=&R_{22}\sin ^{2}\theta .  \label{schw}
\end{eqnarray}%
From this we have that the scalar curvature 
\begin{eqnarray}
R &=&g^{\mu \nu }R_{\mu \nu }=-\left( -\nu ^{\prime \prime }+\lambda
^{\prime }\nu ^{\prime }-\nu ^{\prime 2}-\frac{2\nu ^{\prime }}{r}\right)
e^{-2\lambda }+\left( \nu ^{\prime \prime }-\lambda ^{\prime }\nu ^{\prime
}+\nu ^{\prime 2}-\frac{2\lambda ^{\prime }}{r}\right) e^{-2\lambda }  \notag
\\
&&+\frac{2(1+r\nu ^{\prime }-r\lambda ^{\prime })e^{-2\lambda }-2}{r^{2}} 
\notag \\
&=&\left( 2\nu ^{\prime \prime }-2\lambda ^{\prime }\nu ^{\prime }+2\nu
^{\prime 2}-\frac{4\lambda ^{\prime }-4\nu ^{\prime }}{r}+\frac{2}{r^{2}}%
\right) e^{-2\lambda }-\frac{2}{r^{2}}
\end{eqnarray}

For our model for $T_{\mu \nu }$ \ we take that of a perfect fluid with
pressure $p$, density $\rho $, and four velocity $u^{\mu }$\cite{wein} 
\begin{eqnarray}
T_{\mu \nu } &=&pg_{\mu \nu }+(p+\rho )u_{\mu }u_{\nu },  \notag \\
g_{\mu \nu }u^{\mu }u^{\nu } &=&-1.
\end{eqnarray}%
We work in the frame in which the fluid is at rest, $\mathbf{u=0,}$ and so%
\begin{eqnarray}
g_{00}u^{02} &=&-1,  \notag \\
u_{0} &=&g_{00}u^{0}=g_{00}(-g_{00})^{-1/2},  \notag \\
u_{0}^{2} &=&-g_{00}.
\end{eqnarray}%
Thus, the only nonzero elements of $T_{\mu \nu }$ are 
\begin{eqnarray}
T_{00} &=&pg_{00}-g_{00}(p+\rho )=-g_{00}\rho ,  \notag \\
T_{11} &=&pg_{11},  \notag \\
T_{22} &=&pg_{22},  \label{pf} \\
T_{33} &=&pg_{33}.  \notag
\end{eqnarray}%
The Einstein equations%
\begin{equation}
G_{\mu \nu }=-\kappa T_{\mu \nu }  \label{equ}
\end{equation}
now become%
\begin{eqnarray}
G_{00} &=&R_{00}-\frac{1}{2}g_{00}R=\kappa g_{00}\rho  \notag \\
G_{00} &=&\left( -\nu ^{\prime \prime }+\lambda ^{\prime }\nu ^{\prime }-\nu
^{\prime 2}-\frac{2\nu ^{\prime }}{r}\right) e^{2\nu -2\lambda }+\left( \nu
^{\prime \prime }-\lambda ^{\prime }\nu ^{\prime }+\nu ^{\prime 2}-\frac{%
2\lambda ^{\prime }-2\nu ^{\prime }}{r}+\frac{1}{r^{2}}\right) e^{2\nu
-2\lambda }-\frac{e^{2\nu }}{r^{2}}  \notag \\
&=&\left( -\frac{2\lambda ^{\prime }}{r}+\frac{1}{r^{2}}\right) e^{2\nu
-2\lambda }-\frac{e^{2\nu }}{r^{2}}=-\kappa e^{2\nu }\rho ,  \notag \\
-\kappa \rho &=&e^{-2\lambda }\left( -\frac{2\lambda ^{\prime }}{r}+\frac{1}{%
r^{2}}\right) -\frac{1}{r^{2}},
\end{eqnarray}%
and%
\begin{eqnarray}
G_{11} &=&R_{11}-\frac{1}{2}g_{11}R=-\kappa pg_{11},  \notag \\
-\kappa pe^{2\lambda } &=&\nu ^{\prime \prime }-\lambda ^{\prime }\nu
^{\prime }+\nu ^{\prime 2}-\frac{2\lambda ^{\prime }}{r}-\frac{1}{2}%
e^{2\lambda }(\left( 2\nu ^{\prime \prime }-2\lambda ^{\prime }\nu ^{\prime
}+2\nu ^{\prime 2}-\frac{4\lambda ^{\prime }-4\nu ^{\prime }}{r}+\frac{2}{%
r^{2}}\right) e^{-2\lambda }-\frac{2}{r^{2}})  \notag \\
&=&-\frac{1}{2}e^{2\lambda }(\left( +\frac{4\nu ^{\prime }}{r}+\frac{2}{r^{2}%
}\right) e^{-2\lambda }-\frac{2}{r^{2}}),  \notag \\
-\kappa p &=&e^{-2\lambda }\left( -\frac{2\nu ^{\prime }}{r}-\frac{1}{r^{2}}%
\right) +\frac{1}{r^{2}},
\end{eqnarray}%
and%
\begin{eqnarray}
G_{22} &=&R_{22}-\frac{1}{2}g_{22}R=-\kappa pg_{22},  \notag \\
-\kappa pr^{2} &=&(1+r\nu ^{\prime }-r\lambda ^{\prime })e^{-2\lambda }-1-%
\frac{1}{2}r^{2}(\left( 2\nu ^{\prime \prime }-2\lambda ^{\prime }\nu
^{\prime }+2\nu ^{\prime 2}-\frac{4\lambda ^{\prime }-4\nu ^{\prime }}{r}+%
\frac{2}{r^{2}}\right) e^{-2\lambda }-\frac{2}{r^{2}})  \notag \\
-\kappa p &=&(\frac{\nu ^{\prime }}{r}-\frac{\lambda ^{\prime }}{r}%
)e^{-2\lambda }-(\left( \nu ^{\prime \prime }-\lambda ^{\prime }\nu ^{\prime
}+\nu ^{\prime 2}-\frac{2\lambda ^{\prime }-2\nu ^{\prime }}{r}\right)
e^{-2\lambda })  \notag \\
&=&-\left( \nu ^{\prime \prime }-\lambda ^{\prime }\nu ^{\prime }+\nu
^{\prime 2}-\frac{\lambda ^{\prime }-\nu ^{\prime }}{r}\right) e^{-2\lambda
},
\end{eqnarray}%
and the fourth equation, the one for $G_{33},$ gives nothing new beyond that
for $G_{22}$. \ Hence, the above three simultaneous equations become%
\begin{eqnarray}
-\kappa \rho &=&e^{-2\lambda }\left( -\frac{2\lambda ^{\prime }}{r}+\frac{1}{%
r^{2}}\right) -\frac{1}{r^{2}},  \notag \\
-\kappa p &=&e^{-2\lambda }\left( -\frac{2\nu ^{\prime }}{r}-\frac{1}{r^{2}}%
\right) +\frac{1}{r^{2}},  \notag \\
-\kappa p &=&-\left( \nu ^{\prime \prime }-\lambda ^{\prime }\nu ^{\prime
}+\nu ^{\prime 2}-\frac{\lambda ^{\prime }-\nu ^{\prime }}{r}\right)
e^{-2\lambda }.  \label{ein}
\end{eqnarray}%
These are three nonlinear inhomogeneous equations for two unknown functions (%
$\lambda $ and $\nu $) of $r$.

For empty space ($\rho =p=0)$ these equations become those originally solved
by Schwarzschild, that is.%
\begin{eqnarray}
e^{-2\lambda }\left( -\frac{2\lambda ^{\prime }}{r}+\frac{1}{r^{2}}\right) -%
\frac{1}{r^{2}} &=&0  \notag \\
e^{-2\lambda }\left( -\frac{2\nu ^{\prime }}{r}-\frac{1}{r^{2}}\right) +%
\frac{1}{r^{2}} &=&0,  \notag \\
-\left( \nu ^{\prime \prime }-\lambda ^{\prime }\nu ^{\prime }+\nu ^{\prime
2}-\frac{\lambda ^{\prime }-\nu ^{\prime }}{r}\right) e^{-2\lambda } &=&0.
\end{eqnarray}%
Combining the first two equations implies that 
\begin{eqnarray}
\lambda ^{\prime } &=&-\nu ^{\prime },  \notag \\
\lambda &=&-\nu +\lambda _{0}(t).
\end{eqnarray}%
The third equation then yields%
\begin{equation}
\nu ^{\prime \prime }+2\nu ^{\prime 2}+\frac{2\nu ^{\prime }}{r}=0.
\label{homo}
\end{equation}

We parametrize the exponential metric function $\nu $ by introducing the
potential-like function $\phi $ ,%
\begin{eqnarray}
\nu &=&\frac{1}{2}\ln (1+2\phi ),  \notag \\
\nu ^{\prime } &=&\phi ^{\prime }\frac{1}{1+2\phi }  \notag \\
\nu ^{\prime \prime } &=&\phi ^{\prime \prime }\frac{1}{1+2\phi }-2\phi
^{\prime 2}\frac{1}{\left( 1+2\phi \right) ^{2}}=\phi ^{\prime \prime }\frac{%
1}{1+2\phi }-2\nu ^{\prime 2}.  \label{nuv}
\end{eqnarray}%
Thus, using 
\begin{equation*}
e^{2\nu }=(1+2\phi ),
\end{equation*}%
Eq. (\ref{homo}) becomes%
\begin{equation}
\phi ^{\prime \prime }+\frac{2\phi ^{\prime }}{r}=0.  \label{one}
\end{equation}%
This linear second order homogeneous equation is an equidimensional one and
has the general solution of%
\begin{equation}
\phi =\frac{k_{1}}{r}+k_{2}.
\end{equation}%
Our metric is thus%
\begin{eqnarray}
e^{2\nu } &=&1+\frac{2k_{1}}{r}+2k_{2}=-g_{00}  \notag \\
e^{2\lambda } &=&e^{-2\nu +2\lambda _{0}}=\frac{e^{+2\lambda _{0}}}{1+\frac{%
2k_{1}}{r}+2k_{2}}.
\end{eqnarray}%
In order for the metric to become Minkowskian at $r\rightarrow \infty $ we
must have%
\begin{equation}
k_{2}=0=\lambda _{0}.
\end{equation}%
Matching $g_{00}$ to $-1-2\phi $ for large $r$ where $\phi $ is the
Newtonian potential $-MG/r$ gives 
\begin{equation}
k_{1}=-MG,
\end{equation}%
the Schwarzschild radius, and hence the usual Schwarzschild solution of 
\begin{eqnarray}
g_{00} &=&-1-2\phi =-1+\frac{2MG}{r},  \notag \\
g_{11} &=&1/(1-2MG/r),
\end{eqnarray}%
with the remaining components the same as for free space.

\ The other set of exact solutions and that which is the focus of this paper
is found by assuming that\footnote{%
\ In \cite{deser} a source term for a perfect fluid but with no pressure
term \ is considered. They demonstrate that the point charge is a completely
stable object, without any ad hoc pressure terms required, and its mass is
completely determined by its field interactions.} 
\begin{equation}
p=-\rho \neq 0,  \label{rp}
\end{equation}%
so that Eq. (\ref{pf}) gives us%
\begin{equation}
T_{\mu \nu }=-\rho g_{\mu \nu },
\end{equation}%
and the Einstein equation becomes%
\begin{equation}
G_{\mu \nu }=-\kappa T_{\mu \nu }=\kappa \rho g_{\mu \nu }.  \label{eeq}
\end{equation}%
Just as with the Schwarzschild solution with $\rho =p=0,$ combining the
first two equations of (\ref{ein}) implies that 
\begin{eqnarray}
\nu ^{\prime } &=&-\lambda ^{\prime },  \notag \\
\nu &=&-\lambda +\nu _{0}(t).
\end{eqnarray}%
In this case we absorb the factor $\nu _{0}(t)$ into a redefinition of the
time scale used in the metric. \ Thus we have $\nu =-\lambda $ and the last
two equations of (\ref{ein}) become%
\begin{eqnarray}
-\kappa p &=&-e^{2\nu }\left( \frac{2\nu ^{\prime }}{r}+\frac{1}{r^{2}}%
\right) +\frac{1}{r^{2}}  \notag \\
-\kappa p &=&-\left( \nu ^{\prime \prime }+2\nu ^{\prime 2}+\frac{2\nu
^{\prime }}{r}\right) e^{2\nu }.  \label{cru}
\end{eqnarray}%
Note that these two equations determine both the metric function $\nu (r)$\
and the pressure (and thus the density) so that one does not have a freedom
of choice for the pressure and density beyond Eq. (\ref{rp}).

We parametrize the exponential metric function $\nu $ by introducing a
potential-like function $\phi $ just as in Eq. (\ref{nuv}) so that the last
of the two crucial Einstein equations in (\ref{cru}) become%
\begin{equation}
-\kappa p=-\left( \nu ^{\prime \prime }+2\nu ^{\prime 2}+\frac{2\nu ^{\prime
}}{r}\right) e^{2\nu }=\phi ^{\prime \prime }+\frac{2\phi ^{\prime }}{r}.
\label{pt}
\end{equation}%
Substituting this in the first of Eqs. (\ref{cru}) we obtain 
\begin{eqnarray}
\phi ^{\prime \prime }+\frac{2\phi ^{\prime }}{r} &=&-\kappa p=-(1-2\phi
)\left( -\frac{2\phi ^{\prime }}{(1-2\phi )r}+\frac{1}{r^{2}}\right) +\frac{1%
}{r^{2}}  \notag \\
&=&\frac{2\phi ^{\prime }}{r}+\frac{2\phi }{r^{2}}
\end{eqnarray}%
This leads to the second order linear homogeneous equation%
\begin{equation}
\phi ^{\prime \prime }=\frac{2\phi }{r^{2}}.
\end{equation}%
Note the difference between this equidimensional equation and (\ref{one})
for the homogeneous case. \ Thus one has the general solution of a linear
combination of a harmonic oscillator (for positive $k$) with the Newtonian
potential, 
\begin{equation}
\phi (r)=\frac{1}{2}kr^{2}-\frac{MG}{r}.
\end{equation}%
\ Our metric is thus 
\begin{eqnarray}
g_{00} &=&-e^{2\nu }=-1+kr^{2}-2MG/r=1/g^{00}  \notag \\
g_{11} &=&e^{2\lambda }=e^{-2\nu }=1/(1+kr^{2}-2MG/r)=1/g^{11},  \notag \\
g_{22} &=&r^{2}=1/g^{22},  \notag \\
g_{33} &=&r^{2}\sin ^{2}\theta =1/g^{33}.  \notag \\
g_{\mu \nu } &=&0,~\mu \neq \nu .
\end{eqnarray}%
This corresponds to what is called the Schwarzschild de Sitter space \cite%
{rin}. Without the Newtonian term it corresponds to the solution obtained by
de Sitter for the Einstein equation with a cosmological constant\cite{desit2}%
,\cite{desit4} if $\rho $ is a constant.. \ 

\ As it turns out, the Einstein equations for a perfect fluid plus the
equation of state $p=-\rho $ with no assumption about their space-time
dependence \textit{requires} them to be constants. \ The value of the
constant is determined by Eq. (\ref{pt}) 
\begin{eqnarray}
-\kappa p &=&(\phi ^{\prime \prime }+\frac{2\phi ^{\prime }}{r})=3k,  \notag
\\
\rho &=&-p=-\frac{3k}{\kappa }.  \label{shn}
\end{eqnarray}%
\ The contributions to the pressure and density from the Newtonian part of
the potential vanishes. \ Note because of this determination that $\rho $
and $p$ from the Einstein equation \ref{equ} be constant, the use of Eqs. (%
\ref{eeq}) and (\ref{pf}) implies that this is equivalent to starting with
the \ Einstein equation with a cosmological \textit{constant} $\Lambda $ and
no source term 
\begin{equation}
G_{\mu \nu }+\Lambda g_{\mu \nu }=0,
\end{equation}%
where $\Lambda =-\kappa \rho =3k,$ so that a positive cosmological constant
corresponding to a negative density \footnote{%
Normally one starts with a cosmological constant and then shows the
equivalence to a solution of the ordinary Einstein equation in the presence
of a perfect fluid with $\rho =-p=-\frac{\Lambda }{\kappa }$ \cite{rid}$.$ \
In this paper, we start with $\rho =-p$ with no assumption about their space
time dependence and show that the Einstein equations then force them to be
space-time independent. \ This is a subtle difference not emphasized in most
text books. Unlike starting with a cosmological constant where the first of
Eqs. ( \ref{cru}) would be (as in \cite{rin}), $-1+\Lambda r^{2}=-e^{2\nu
}\left( 2r\nu ^{\prime }+1\right) $\ which can be readily solved, in the
approach given here that first equation cannot be solved since $p$ is an
unknown. Instead one must use the second of Eqs. (\ref{cru}) to eliminate $p$%
, solve directly for $V$ and \textit{then} determine $p$ from that solution.}%
.

This vanishing of the source term for the Newtonian portion of the metric is
contrary to what is expected based on what occurs in the Poisson equation
for a point mass density%
\begin{eqnarray}
\nabla ^{2}\Phi &=&4\pi GM\delta ^{3}(\mathbf{r)=}4\pi G\varrho \mathbf{,} 
\notag \\
\Phi &=&-\frac{GM}{r}.  \label{po}
\end{eqnarray}%
\ Is there a point mass at the origin in the case of the Einstein equation?
This would seem to be implied by the above Newtonian-Poisson connection. \
One may be tempted to replace $\phi ^{\prime \prime }+\frac{2\phi ^{\prime }%
}{r}$ with $\nabla ^{2}\phi $ and proclaim that $-\kappa \rho \equiv -\kappa
\varrho /2=\kappa p=-\nabla ^{2}\phi =-4\pi MG\delta ^{3}(\mathbf{r)-}3k$
but for regions that do not exclude the origin (in contrast see Eq. (\ref%
{exc}) below) this is not consistent with the other expression for the
pressure of $-\kappa \rho \equiv -\kappa \varrho /2=\kappa p=-2\phi
~^{\prime }/r-2\phi /r^{2}=-3k~$\footnote{%
The reason for the introduction of $\varrho \equiv 2\rho $ is that in the
limit, $-g_{00}=1-2\phi \rightarrow 1-2\Phi $ and so we would have $-\kappa
\rho =-8\pi G\rho =-\nabla ^{2}\Phi $ which disagrees with the ordinary
source term in the Poisson equation by a factor of $2.$\ Thus, we take $\rho
\equiv \varrho /2.$}

. \ This calls for a more careful treatment of the Einstein equation near
the origin. \ 

\subsection{The Schwarzschild Solution as a Valid Approximation for a
Nonlinear Solution of the Full Non-homogeneous Einstein Equation}

\ In order to treat the problem at the origin more carefully and achieve the
aim of this paper, we view the Einstein equations in their reduced forms
given in (\ref{cru}) as the limit for small $\varepsilon $ of the following
pair 
\begin{eqnarray}
-\kappa p &=&-e^{2\nu }\left( \frac{2\nu ^{\prime }}{\bar{r}}+\frac{1}{\bar{r%
}^{2}}\right) +\frac{1}{\bar{r}^{2}},  \notag \\
-\kappa p &=&-\left( \nu ^{\prime \prime }+2\nu ^{\prime 2}+\frac{2\nu
^{\prime }}{\bar{r}}\right) e^{2\nu }.
\end{eqnarray}%
where 
\begin{equation}
\bar{r}\equiv \left( r^{2}+\varepsilon ^{2}\right) ^{1/2}.
\end{equation}%
We shall solve these equations instead of the reduced forms (\ref{cru}) of
the Einstein equations and view the proper solutions of the Einstein
equation as the limit of small $\varepsilon $ of the solutions of the
modified equations. Thus, as before, using Eq. (\ref{nuv}) the two crucial
Einstein equations in (\ref{cru}) become%
\begin{eqnarray}
-\kappa p &=&\kappa \rho =(\phi ^{\prime \prime }+\frac{2\phi ^{\prime }}{%
\bar{r}})=-e^{2\nu }\left( -\frac{2\phi ^{\prime }}{\left( 1-2\phi \right) 
\bar{r}}+\frac{1}{\bar{r}^{2}}\right) +\frac{1}{\bar{r}^{2}}  \notag \\
&=&\frac{2\phi ^{\prime }}{\bar{r}}+\frac{2\phi }{\bar{r}^{2}},  \label{36}
\end{eqnarray}%
and this leads to%
\begin{equation}
\phi ^{\prime \prime }=\frac{2\phi }{\bar{r}^{2}}=\frac{2\phi }{%
(r^{2}+\varepsilon ^{2})}.  \label{rbr}
\end{equation}%
Clearly, one solution is 
\begin{equation}
\phi _{1}(r,\varepsilon )=\frac{\kappa _{1}}{2}(r^{2}+\varepsilon ^{2}).
\end{equation}%
Using the connection%
\begin{equation}
\phi _{2}=\kappa _{2}\phi _{1}\int \frac{dr}{\phi _{1}^{2}},
\end{equation}%
between the first and second solution of a homogeneous second order
differential equation, the second solution is 
\begin{equation}
\phi _{2}(r,\varepsilon )=\frac{2\kappa _{2}(r^{2}+\varepsilon ^{2})}{\kappa
_{1}}\int^{r}\frac{dr^{\prime }}{(r^{\prime 2}+\varepsilon ^{2})^{2}},
\end{equation}%
in which both $\kappa _{1}$ and $\kappa _{2}$ are constants. Redefine $%
\kappa _{2}/\kappa _{1}$ as $\kappa _{2}$ and choose the lower limit to be $%
r=\infty $ so that, performing the integration,

\begin{eqnarray}
\phi _{2}(r,\varepsilon ) &=&\frac{\kappa _{2}(r^{2}+\varepsilon ^{2})}{%
\varepsilon }[\frac{1}{\varepsilon ^{2}}\left( \arctan \frac{r}{\varepsilon }%
-\frac{\pi }{2}\right) +\frac{r}{\varepsilon \left( r^{2}+\varepsilon
^{2}\right) }]  \label{v2} \\
&=&\frac{\kappa _{2}(r^{2}+\varepsilon ^{2})}{\varepsilon }[-\frac{1}{%
\varepsilon ^{2}}\left( \arctan \frac{\varepsilon }{r}\right) +\frac{r}{%
\varepsilon \left( r^{2}+\varepsilon ^{2}\right) }]  \notag
\end{eqnarray}

If we let $\varepsilon \rightarrow 0,$ we should get (to match with the
Newtonian solution for large $r$%
\begin{equation}
\phi _{2}(r,0)=-2\kappa _{2}r^{2}\int_{r}^{\infty }\frac{dr^{\prime }}{%
r^{\prime 4}}=-\frac{2\kappa _{2}r^{2}}{3r^{3}}=-\frac{2\kappa _{2}}{3r}=-%
\frac{MG}{r},
\end{equation}%
so we take%
\begin{equation}
\kappa _{2}=\frac{3MG}{2}.
\end{equation}%
Let us check that with this choice our integrated result (\ref{v2}) has this
same limit 
\begin{eqnarray*}
&&\frac{3MG}{2}\frac{(r^{2}+\varepsilon ^{2})}{\varepsilon }[-\frac{1}{%
\varepsilon ^{2}}\left( \arctan \frac{\varepsilon }{r}\right) +\frac{r}{%
\varepsilon \left( r^{2}+\varepsilon ^{2}\right) }] \\
&\rightarrow &\frac{3MG}{2}\frac{(r^{2}+\varepsilon ^{2})}{\varepsilon }[-%
\frac{1}{\varepsilon ^{2}}\left( \frac{\varepsilon }{r}-\frac{1}{3}\left( 
\frac{\varepsilon }{r}\right) ^{3}\right) +\frac{1}{\varepsilon r\left(
1+\varepsilon ^{2}/r^{2}\right) }] \\
&\rightarrow &\frac{3MG}{2}\frac{(r^{2})}{\varepsilon }[-\frac{1}{%
\varepsilon ^{2}}\left( \frac{\varepsilon }{r}-\frac{1}{3}\left( \frac{%
\varepsilon }{r}\right) ^{3}\right) +\frac{1}{\varepsilon r}-\frac{%
\varepsilon }{r^{3}}] \\
&=&\frac{3MG}{2}\frac{(r^{2})}{\varepsilon }\left[ -\frac{2\varepsilon }{%
3r^{3}}\right] =-\frac{MG}{r}
\end{eqnarray*}%
Thus, our general solution to Eq. (\ref{rbr}) is%
\begin{equation}
\phi (r,\varepsilon )=.\frac{\kappa _{1}}{2}(r^{2}+\varepsilon ^{2})+\frac{%
3MG}{2}\frac{(r^{2}+\varepsilon ^{2})}{\varepsilon ^{3}}[\frac{\varepsilon r%
}{(r^{2}+\varepsilon ^{2})}-\arctan \varepsilon /r]  \label{gns}
\end{equation}%
with the corresponding metric given by%
\begin{eqnarray}
g_{00} &=&-e^{2\nu }=-1-2\phi (r,\varepsilon )=1/g^{00}  \notag \\
g_{11} &=&e^{2\lambda }=e^{-2\nu }=\frac{1}{1+2\phi (r,\varepsilon )}%
=1/g^{11},  \notag \\
g_{22} &=&r^{2}=1/g^{22},  \notag \\
g_{33} &=&r^{2}\sin ^{2}\theta =1/g^{33},  \notag \\
g_{\mu \nu } &=&0,~\mu \neq \nu .
\end{eqnarray}

Now, let us determine what the density and pressure are for the limit of
small $\varepsilon $. This will provide us with insight into the nature of
the source term.\ The simplest way is to evaluate $\kappa \rho \equiv \kappa
\varrho /2=2\phi ^{\prime }/\bar{r}+2\phi /\bar{r}^{2}$. \ We obtain

\begin{eqnarray}
&&\kappa \varrho (r,\varepsilon )/2  \notag \\
&=&\kappa \lbrack \varrho _{1}(r,\varepsilon )/2+\varrho _{2}(r,\varepsilon
)/2]  \notag \\
&=&+\kappa _{1}(\frac{2r}{(r^{2}+\varepsilon ^{2})^{1/2}}+1)  \notag \\
&&+\frac{6MG}{\varepsilon ^{2}(r^{2}+\varepsilon ^{2})^{1/2}}[1-\frac{r}{%
\varepsilon }\arctan \varepsilon /r]+\frac{6MG}{2\varepsilon ^{3}}[\frac{%
\varepsilon r}{(r^{2}+\varepsilon ^{2})}-\arctan \varepsilon /r],  \label{52}
\end{eqnarray}%
where $\varrho _{1}$ \ is the density that arises from the oscillator-like
part of the solution while $\varrho _{2}$ \ is the density that arises from
the Newtonian-like part of the solution. For $r\neq 0$ and $\varepsilon
\rightarrow 0$ 
\begin{equation}
\kappa \varrho _{2}/2\rightarrow \frac{6MG}{\varepsilon ^{2}r}[\frac{%
\varepsilon ^{2}}{3r^{2}}]+\frac{6MG}{2\varepsilon ^{3}}[\frac{\varepsilon }{%
r}-\frac{\varepsilon ^{3}}{r^{3}}-\frac{\varepsilon }{r}+\frac{\varepsilon
^{3}}{3r^{3}}]~~=\frac{6MG}{\varepsilon ^{2}r}[\frac{\varepsilon ^{2}}{3r^{2}%
}]+\frac{6MG}{2\varepsilon ^{3}}[-\frac{2\varepsilon ^{3}}{3r^{3}}]=0,
\end{equation}%
while for $r=0$ and $\varepsilon \rightarrow 0$ this becomes 
\begin{equation}
\kappa \varrho _{2}/2\rightarrow \frac{6MG}{\varepsilon ^{3}}(1-\frac{\pi }{4%
})
\end{equation}%
These two limits taken together have the appearance of a delta function
expected for a point mass. \ To complete the verification let us check by
integrating the Newtonian term for the density over a sphere of radius $R$
and using the divergence theorem, that we obtain the appropriate constant
(independent of $\varepsilon $). From (\ref{36}) and (\ref{rbr}) we find that

\begin{eqnarray}
&&\kappa \int d^{3}r\varrho _{2}(r,\varepsilon )/2  \notag \\
&=&6MG\int d^{3}r[\frac{1}{\varepsilon ^{2}(r^{2}+\varepsilon ^{2})^{1/2}}[1-%
\frac{r}{\varepsilon }\arctan \varepsilon /r]+\frac{1}{2\varepsilon ^{3}}[%
\frac{\varepsilon r}{(r^{2}+\varepsilon ^{2})}-\arctan \varepsilon /r]] 
\notag \\
&=&\int d^{3}r[\frac{2\phi ^{\prime }}{\bar{r}}+\frac{2\phi }{\bar{r}^{2}}%
]=\int d^{3}r(\phi ^{\prime \prime }+\frac{2\phi ^{\prime }}{\bar{r}}).
\end{eqnarray}%
For small $\varepsilon $ $\left( \varepsilon /R\rightarrow 0\right) $ this
integral becomes 
\begin{eqnarray}
&\rightarrow &\int d^{3}r(\phi ^{\prime \prime }+\frac{2\phi ^{\prime }}{%
\bar{r}})=\int d^{3}r\nabla ^{2}\phi =\underset{\varepsilon /R\rightarrow 0}{%
\lim }R^{2}4\pi \phi ^{\prime }(R)  \notag \\
&=&4\pi \underset{\varepsilon /R\rightarrow 0}{\lim }R^{2}3MG\left[ \frac{1}{%
\varepsilon ^{2}}-\frac{R}{\varepsilon ^{3}}\arctan \varepsilon /R\right]
=12\pi MG\underset{\varepsilon /R\rightarrow \infty }{\lim }\frac{R^{3}}{%
\varepsilon ^{3}}(\frac{\varepsilon }{R}-\frac{\varepsilon }{R}+\frac{1}{3}%
\left( \frac{\varepsilon }{R}\right) ^{3})  \notag \\
&=&4\pi GM=4\pi G\int d^{3}r\varrho _{2}(r,\varepsilon )
\end{eqnarray}%
Such would not be the case for the homogeneous equation which would give
zero for the integrated density (see below Eq. (\ref{shn})). \ Thus, in the
limit $\varepsilon \rightarrow 0$ where our $\varepsilon -$modified Einstein
equations become the actual Einstein equation, the integral of the density
over an arbitrarily small volume remains a constant independent of $%
\varepsilon $. \ \ So, rearranging the terms in $\rho _{2}$, we define%
\begin{eqnarray}
\delta ^{3}(\mathbf{r,}\varepsilon ) &\equiv &\frac{\kappa \varrho
_{2}(r,\varepsilon )/2}{4\pi GM}=\frac{3}{2\pi \varepsilon ^{3}}[\frac{%
\varepsilon /r}{(1+\left( \varepsilon /r\right) ^{2})^{1/2}}(1+\frac{1}{2}%
\frac{1}{(1+\left( \varepsilon /r\right) ^{2})^{1/2}})  \notag \\
&&-\arctan \varepsilon /r(\frac{1}{(1+\left( \varepsilon /r\right)
^{2})^{1/2}}+\frac{1}{2})],
\end{eqnarray}%
with the property that%
\begin{equation}
\int d^{3}r\delta ^{3}(\mathbf{r,}\varepsilon )=1.
\end{equation}%
\ Our $\delta ^{3}(\mathbf{r,}\varepsilon )$ therefore has the requisite
properties for a distribution that in the limit represents a Dirac delta
function. \ Its value for $\mathbf{r\neq 0}$ tends to zero as $\varepsilon
\rightarrow 0$ and its integral over all space is unity. \ This establishes
that for $\kappa _{1}=0,$ the source term is non-zero and has the property
of a sharply confined distribution. \ What makes this source distinct from
others \cite{delta}, which begin with the matching solution of Schwarzschild 
\cite{sch} to an incompressible fluid confined within a finite spherical
surface, is that within this sharply confined region, $p=-\rho =\varrho /2$.%
\footnote{%
In \cite{deser} a different approach also leads to a delta function like
source term for the Einstein equation for static, spherically symmetric
circumstances. \ They assume a dust with no pressure term present, confined
in a radius $\varepsilon $ composed of charged particles. \ They demonstrate
that only in the $\varepsilon \rightarrow 0$ limit is the solution stable
and static, corresponding to a charged point particles. That is, in that
limit and only in that limit, they find that\ gravitational forces have
exactly counteracted the repulsive electrostatic self-forces. \ }

The equations analogous to Eq. (\ref{po}) are given in Eq. (\ref{36}), in
particular%
\begin{eqnarray}
\kappa \rho &=&\kappa \varrho /2=\phi ^{\prime \prime }+\frac{2\phi ^{\prime
}}{\bar{r}}=\nabla ^{2}\phi (r,\varepsilon )  \notag \\
&=&4\pi G\varrho =4\pi GM\delta ^{3}(\mathbf{r,}\varepsilon ),  \notag \\
\phi (r,\varepsilon ) &=&.\frac{\kappa _{1}}{2}(r^{2}+\varepsilon ^{2})+%
\frac{3MG}{2}\frac{(r^{2}+\varepsilon ^{2})}{\varepsilon ^{3}}[\frac{%
\varepsilon r}{(r^{2}+\varepsilon ^{2})}-\arctan \varepsilon /r]  \notag \\
&\rightarrow &\frac{\kappa _{1}r^{2}}{2}-\frac{GM}{r}
\end{eqnarray}

\subsubsection{The Schwarzschild Limit}

We consider in this section the potential-like function $\phi (r,\varepsilon
)$ given in Eq. (\ref{gns}). \ We wish to determine in what sense that, if
we choose $\kappa _{1}=0,$ the second portion of $\phi (r,\varepsilon )$ for 
$\varepsilon >0$ sufficiently small agrees with the Schwarzschild solution $%
\phi _{s}(r)=-GM/r$ for a given range of $r.$ \ Let us make this statement
precise. \ We show (with $r_{s}=2MG)$ that for a positive $\delta >0$ that
if,$,$%
\begin{equation}
\frac{\varepsilon }{r_{s}}\equiv \epsilon <\delta .
\end{equation}%
then, 
\begin{equation}
\left( \frac{r}{r_{s}}\right) ^{3}\left\vert \phi (r,\varepsilon )-(-)\frac{%
GM}{r}\right\vert <\frac{\delta ^{2}}{10}.  \label{ineq}
\end{equation}%
for $r$ in the range $\varepsilon <r<\infty $. \ If $\delta <1$, then the
range for $r$ of agreement between the potentials in the above sense would
extend down below the Schwarzschild radius with\ no upper bound. \ 

To show this we consider the Taylor series for $\phi (r,\varepsilon )-\phi
_{s}(r)$ in $\varepsilon $ about $\varepsilon =0.$ 
\begin{equation}
\phi (r,\varepsilon )-(-)\frac{GM}{r}=\frac{3MG}{2}\frac{(r^{2}+\varepsilon
^{2})}{\varepsilon ^{3}}[\frac{\varepsilon r}{(r^{2}+\varepsilon ^{2})}%
-\arctan \varepsilon /r]+\frac{GM}{r}
\end{equation}%
The series for $\arctan \varepsilon /r$ converges for $r>\varepsilon $. \
Expanding we find that%
\begin{eqnarray}
\phi (r,\varepsilon )-(-)\frac{GM}{r} &=&\frac{3MG}{2\varepsilon ^{3}}%
[\varepsilon r-r^{2}(1+\frac{\varepsilon ^{2}}{r^{2}})\sum_{n=0}^{\infty }%
\frac{(-)^{n}}{2n+1}\left( \frac{\varepsilon }{r}\right) ^{2n+1}]+\frac{GM}{r%
}  \notag \\
&=&\frac{3MG}{2\varepsilon ^{3}}\left[ -2\varepsilon ^{2}\sum_{n=0}^{\infty
}\left( \frac{1}{\left( 2n+3\right) \left( 2n+1\right) }\right)
(-)^{n}\left( \frac{\varepsilon }{r}\right) ^{2n+1}\right] +\frac{GM}{r} 
\notag \\
&=&-\frac{3MG}{\varepsilon }[\left( \frac{\varepsilon }{r}\right)
(\sum_{n=0}^{\infty }(\frac{1}{\left( 2n+3\right) \left( 2n+1\right) }%
)(-)^{n}\left( \frac{\varepsilon }{r}\right) ^{2n}]+\frac{GM}{r}  \notag \\
&=&-3MG[\left( \frac{1}{r}\right) (\frac{1}{3}+\sum_{n=1}^{\infty }(\frac{1}{%
\left( 2n+3\right) \left( 2n+1\right) })(-)^{n}\left( \frac{\varepsilon }{r}%
\right) ^{2n}]+\frac{GM}{r}  \notag \\
&=&-\frac{3MG}{r}[\sum_{n=1}^{\infty }(\frac{1}{\left( 2n+3\right) \left(
2n+1\right) })(-)^{n}\left( \frac{\varepsilon }{r}\right) ^{2n}]  \notag \\
&=&\frac{3MG}{r}[\sum_{n=0}^{\infty }(\frac{1}{\left( 2n+5\right) \left(
2n+3\right) })(-)^{n}\left( \frac{\varepsilon }{r}\right) ^{2n+2}]<\frac{%
MG\varepsilon ^{2}}{5r^{3}}
\end{eqnarray}%
Thus, with $\frac{\varepsilon }{r_{s}}=\epsilon <\delta ,$ we have since $%
\epsilon \frac{r_{s}}{r}<1$%
\begin{eqnarray}
\left( \frac{r}{r_{s}}\right) ^{3}\left\vert \phi (r,\varepsilon )+\frac{GM}{%
r}\right\vert &=&\frac{1}{2}\left( \frac{r}{r_{s}}\right) ^{2}\left\vert 
\frac{r}{GM}\phi (r,\varepsilon )+1\right\vert =\frac{3r^{2}}{r_{s}^{2}}%
[(\sum_{n=0}^{\infty }(\frac{1}{\left( 2n+5\right) \left( 2n+3\right) }%
)(-)^{n}\left( \frac{\varepsilon }{r}\right) ^{2n+2}]  \notag \\
&=&\frac{3\varepsilon ^{2}}{2r_{s}^{2}}[(\sum_{n=0}^{\infty }(\frac{1}{%
\left( 2n+5\right) \left( 2n+3\right) })(-)^{n}\left( \frac{\varepsilon }{r}%
\right) ^{2n}]  \notag \\
&=&\frac{3\epsilon ^{2}}{2}[(\sum_{n=0}^{\infty }(\frac{3}{\left(
2n+5\right) \left( 2n+3\right) })(-)^{n}\left( \epsilon \frac{r_{s}}{r}%
\right) ^{2n}]  \notag \\
&=&\frac{\epsilon ^{2}}{10}\left( 1+\sum_{n=1}^{\infty }(\frac{15}{2\left(
2n+5\right) \left( 2n+3\right) })(-)^{n}\left( \epsilon \frac{r_{s}}{r}%
\right) ^{2n}\right)  \notag \\
&<&\frac{\epsilon ^{2}}{10}<\frac{\delta ^{2}}{10}.
\end{eqnarray}%
So, we have demonstrated how the Schwarzschild solution can be viewed as a
valid approximation for a bona fide nonlinear solution of the full
non-homogeneous Einstein equation.

\section{Discussion}

We have found that the general solution of the Einstein equation for the
special case of $p=-\rho \equiv \varrho /2$ to yield a metric governed by a
linear combination of a Newtonian and simple harmonic oscillator potential
(for positive $k$) , two independent solutions of a linear second order
differential equation for the potential-like function $\phi $ . \ If the
behavior about the origin is not handled carefully, the $\sim r^{2}$
contribution gives a constant density and pressure while the Newtonian
contribution gives rise to no point-like (delta function) source. \ With $%
\kappa _{1}=0,$ this is the usual no source or vacuum solution.\ Handled
more carefully by the method we present in this paper we obtain density and
pressure terms sharply peaked about the origin, with a unit volume integral.

Do our mathematical solutions of the inhomogeneous Einstein equation have
any physical significance? Superimposed on our point-like density is a
density ranging between two constants ($\kappa _{1}$ and $3\kappa _{1};$ see
Eq. (\ref{52})$)$, the source of the harmonic oscillator potential-like
function $\phi (r,\varepsilon )$ which behaves like $\kappa _{1}r^{2}/2$ for
sufficiently small $r.$ \ The static and spherically symmetric metric we
started with is distinct from the standard time-dependent
Friedmann-Lematre-Robertson-Walker (FLRW) metric. \ It is not intended to
relate to the universe as a whole but rather to the field produced by a
single source. \ There is some superficial similarity between our solution
and the so-called dark energy and inflation solutions of modern cosmology
for $\kappa _{1}<0$ since both involve a pressure with an opposite sign of
the density. \ It may be of just academic interest that the potential-like
function $\phi $ we obtain is not only proportional to $1/r$ but also to the
otherwise ubiquitous $r^{2}$ potential correspond to the solution of the
Einstein equation under these circumstances.\ A positive or negative sign of 
$\kappa _{1}$ would give a negative or positive pressure and an attractive
or repulsive force that would increase in magnitude with distance (mimicking
the effects of dark energy). \ \ Note that in nonrelativistic potential
theory a constant density would give rise to an attractive Hooke's law
force. \ However the context is entirely difference. There, the Hooke's law
\ form follows from Gauss' law applied to an inverse square field. \ The $%
r^{2}$ potential like function discussed in this paper is completely
independent of the $1/r$ potential. \ Another factor to point out is that
the functional form of the density or pressure correlated with the $r^{2}$
potential is fixed by our solution to the Einstein equations themselves, it
is not imposed. \ The only imposition we made on the density and pressure is
that they be the negative of one another. \ From a mathematical point of
view there is no distinction between the solutions discussed here for the
inhomogeneous Einstein equation and the one we would have obtained by adding
a term $+\rho g_{\mu \nu }$ to the left hand side of the Einstein equation
and viewing it as an addition to the equation, in analogy to the alternative
explanation of dark energy. \ The difference here is that $\rho $ being a
constant would be an outcome of the modified Einstein equations and not an
imposed functional form. \ 

It was one of Einstein's early goals, although he never succeeded, to
incorporate Mach's principle in his general theory of relativity. It has
been generally regarded that general relativity does not embody Mach's
principle, that is that geometry can exist independent of matter. \ It was
the Schwarzschild solution that seemed to bring this idea its early but
reluctant acceptance. That is, a geometry can arise in the absence of a
source term, from the vacuum. \ Of course, in the practical applications of
the Schwarzschild solution to the precession problem of Mercury and the
bending of light, it was always assumed that looming behind the formal
sourceless equation was a real sun. \ Nevertheless, a possible formal
interpretation has been that a curved space exists without an identifiable
source, thus obviating the need for Mach's principle

Our result has been to replace the Schwarzschild solution to the sourceless
spherically symmetric static environment, which then, as now seems to allow
the existence of non-trivial spacetime curvature in absence of any matter,
with a solution that does not correspond to a sourceless environment but yet
leads nevertheless to a metric that can approach the Schwarzschild with
arbitrary accuracy in an asymptotic way. In doing so, for this particular
case at least, Mach's principle, the idea that geometry emerges as an
interaction between an identifiable matter term and geometry is preserved%
\cite{mach2}.

\appendix\setcounter{equation}{0} \ \renewcommand{\theequation}{%
\Alph{section}.\arabic{equation}}

\begin{acknowledgement}
The author acknowledges helpful correspondences with Professor M. Sachs and
useful suggestions from Professor G. Longhi, Dr. L. Lusanna, Dr. C. Powell,
J. Labello, and S. Rubenstein
\end{acknowledgement}

\end{document}